\documentclass[preprint]{aastex}
\usepackage{apjfonts}
\usepackage{graphicx}
\usepackage{epsfig}
\usepackage{float}
\usepackage{amsmath}
\usepackage{natbib}
\usepackage{float}
\usepackage{amssymb}

\usepackage[usenames]{color}

\shorttitle{}
\shortauthors{}

\begin{document}
\title{The hard X-ray shortages prompted by the clock bursts in GS 1826--238}

\author{Long Ji\altaffilmark{1}, Shu Zhang\altaffilmark{1}, YuPeng
Chen\altaffilmark{1}, ShuangNan Zhang\altaffilmark{1}, Diego F. Torres\altaffilmark{2,3}, Peter
Kretschmar\altaffilmark{4}, Jian Li\altaffilmark{1}}
\altaffiltext{1}{Laboratory for Particle Astrophysics, Institute of High Energy
Physics, Beijing 100049, China}
\altaffiltext{2}{Instituci\'o Catalana de Recerca i Estudis Avan\c cats (ICREA),
08010 Barcelona, Spain}
\altaffiltext{3}{Institute of Space Sciences (IEEC-CSIC), Campus UAB, Torre C5,
2a planta, 08193 Barcelona, Spain}
\altaffiltext{4}{European Space Astronomy Centre (ESA/ESAC), Science Operations
Department, Villanueva de la Ca\~nada (Madrid), Spain}

\begin{abstract}

We report on a study of GS 1826--238 using all available {\it RXTE} observations, concentrating on the behavior of the hard X-rays during type-I  bursts.
We find a hard X-ray shortage at 30--50 keV promoted by the shower of soft X-rays coming from type-I  bursts. This shortage happens with a time delay after the peak of the soft flux of 3.6 $\pm$ 1.2 seconds.
The behavior of hard X-rays during bursts indicates cooling and reheating of the corona, during which a large amount of energy is required. We speculate that this energy originates from the feedback of the type-I  bursts to the accretion process, resulting in a rapid temporary increase of the accretion rate.

\end{abstract}

\section{Introduction}

Low mass X-ray binaries (XRBs) consist of a compact (neutron star or black hole) and a companion star, with the latter transferring  mass and forming an accretion disc, a process during which the gravitational potential energy is converted into radiation,  see, e.g., \cite{Lewin1993,Strohmayer2006,Galloway2008}, and references therein.
The spectrum is usually modeled having two components: a blackbody plus a power law, in which the former represents the emission of the accretion disc, and the latter is attributed to the contribution of corona or jet  \citep{Remillard2006,Markoff2004}.
If the hard X-rays are from synchrotron radiation in a jet, one would likely observe too the radio emissions in the lower energy band.
However according to radio observations of XRBs, the luminosity of the systems having a neutron star companion is systematically lower than of those having a black hole, by a factor of $\sim$ 30, and the relation of the radio and X-ray luminosity is different between neutron stars and black holes. Therefore, the hard X-rays in neutron star systems more likely derive from inverse Comptonization processes in a corona (for details, please see \citet{Migliari2010,Migliari2006}).

However, although the concept of the corona is widely used in current models, its formation remains less understood. Both, evaporation \citep{Meyer1994,Esin1997,Liu2007,Frank2002} and magnetic reconnection \citep{Zhang2000,Mayer2007} models have been proposed. Type-I bursts provide a fresh input of soft-X-ray photons with which the corona electrons could interact, and thus they may become a useful probe.
Type-I bursts are thermal nuclear explosions that are triggered by the unstable burning on the surface of the neutron star and characterized by intense emissions released in a short time (generally, several tens of seconds) in the form of a blackbody \citep{Swank1977}.
In several other atoll sources, namely IGR J17473-2721, Aql X-1, and 4U 1636-536, a shortage of flux in the hard X-ray band during X-ray bursts has been found \citep{Maccarone2003,Chen2013,Ji2013}.
GS 1826--238 is an atoll source, exhibiting substantial number of X-ray bursts.
It was discovered by the Ginga satellite in 1988 \citep{Makino1988} and its type-I  bursts were first detected by BeppoSAX \citep{Ubertini1997}. The shape and spectrum of bursts in GS 1826-238 change little and the persistent flux is relatively stable \citep{Galloway2004, Thompson2005}. These properties imply  relatively smaller systematic errors when investigating the burst influence when compared with bursts in other sources. In addition, the duration of bursts in GS 1826-238 is much longer than those in other sources, resulting in a more significant shortage in the hard X-rays.
Assuming a distance of 6 kpc, this source is always staying in the low hard state with a luminosity of 5$\%- 9\%$ $L_{\rm edd}$, and a  spectrum dominated by a power law \citep{Galloway2008}. What makes this source unique is the remarkable stability of its burst recurrence rate as a function of mass accretion rate over a timescale of many years. GS 1826--238 is also called the `clock-burster'  because of this fact \citep{Ubertini1999a}. Rossi X-Ray Timing Explorer ({\it RXTE}) observations revealed that the source flux intensity steadily increased between 1997-2003, during which the burst rate increased.
The distance to the source was estimated at 4 -- 8 kpc \citep{Barret1995,intZand1999,Kong2000}. Its hard X-ray emissions are usually modeled by a thermal Comptonization with a hot electron cloud (T $\sim$ 20 keV) \citep{Cocchi2010}.

{In this paper, we report on a study using all available {\it RXTE} observations of GS 1826--238, concentrating on the behavior of the hard X-rays during type-I  bursts.}

%\begin{figure}
%\centering
%\includegraphics[width=8.cm]{asm.eps}
%\caption{The green line shows the long-term ASM count rate history of GS 1826-24. One year averages were calculated from 1 day measurements. The time when type-I  bursts occurred were denoted as red lines.}
%\label{asm}
%\end{figure}

%%%%%%%%%%%%%%%%%%%%%%%%%%%%%%%%%%%%%%%%%
\section{Observations and data analysis}

We analyzed the observations of GS 1826--238 carried out by the All Sky Monitor (ASM; Levine et al. 1996) and the Proportional Counter Array (PCA; Jahoda et al. 1996) on board {\it RXTE}. For the latter, only pointing mode data were used, amounting 912 ks. The PCA was made-up of five identical co-aligned proportional counter units (PCUs) that were sensitive to 2--60 keV photons.
%We have only used PCU 2, due to its better calibration.
We have used all active PCUs during bursts for the following analysis and for each burst we have normalized the lightcurve in unit of counts/s/PCU.
We took the Event-mode data when producing the lightcurves, due to its high time resolution.
For the persistent flux, the background was created based on the latest bright source models using the program {\it pcabackest}, which is a standard software in {\it FTOOLS}.
Note that above 30 keV PCA count rates are dominated by the instrument background, which varies substantially during an orbit\citep{Jahoda1996}. But we found that the persistent flux prior to the bursts is consistent with that after bursts, suggesting the background could lead to little influence on the following analysis because of the short duration of bursts.
Detector breakdowns were eliminated, and the dead time was corrected according to the data of standard~1 mode. The spectrum was fitted by {\it XSPEC} 12.7.1, fixing the hydrogen column density to $0.3 \times 10^{22}$ ${{atoms} \mathord{\left/
 {\vphantom {{atoms} {c{m^2}}}} \right.
 \kern-\nulldelimiterspace} {c{m^2}}}$\citep{Thompson2005}.

To distinguish real bursts from fluctuations in the persistent emission, we searched for bursts that presented a peak flux of at least 500 cts/s larger than the persistent flux in the 2--10 keV energy band. As we show in Section 3, the narrow distributions of the persistent and peak fluxes average at $\sim$94 cts/s and $\sim$ 1917 cts/s, respectively. Hence, such a selection criterion is reasonable and effective to search for the bursts.
We eliminated the bursts for which the observations were incomplete in a time period of 400 seconds around the flux peak (150 seconds before and 250 seconds after).
With these cuts, 43 bursts were selected as our sample.

Since the launch of {\it RXTE} at the end of 1995, GS 1826--238 was observed 160 times in pointing mode. Both PCA and ASM observations showed an increasing luminosity  between 1997 and 2003, with the X-ray intensity flattening afterwards. %, see Figure~\ref{asm}.
%
%The 43 bursts adopted in our analysis %are marked as red lines in Figure~\ref{asm}, and they
%happened along different epochs of observations of the source.

\section{Results}

\begin{figure}[t]
\centering
\includegraphics[width=8.5cm]{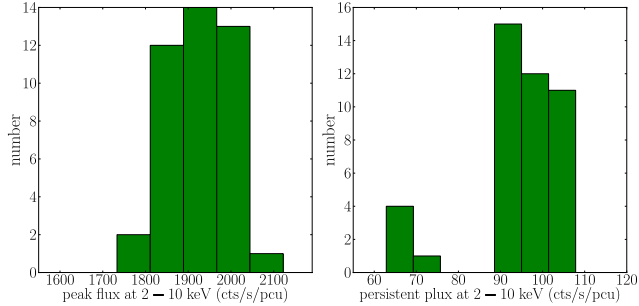}
\caption{The distributions of the persistent and peak fluxes for the type-I  bursts}
\label{peak_and_persistent}
\end{figure}

\begin{figure}[t]
\centering
\includegraphics[width=8.5cm]{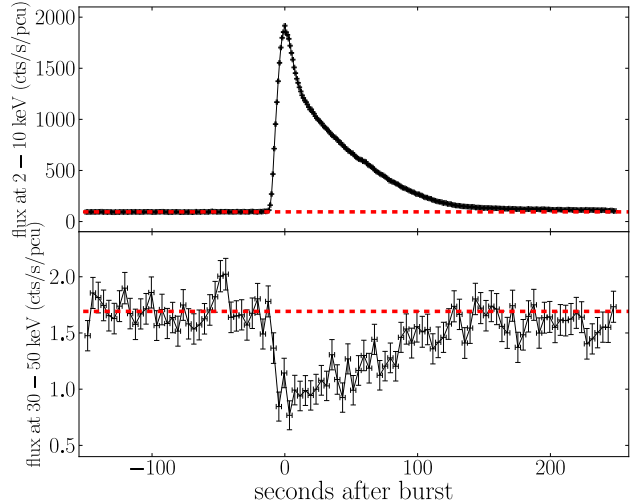}
\caption{The combined lightcurves of all the 43 bursts at energies of 2--10 keV and 30--50 keV.}
\label{total}
\end{figure}

\begin{table*}[t]
\small
\caption{The columns denote the observation ID, the time when type-I  bursts occurred, peak flux of bursts, soft and hard X-ray persistent flux, and the burst duration.}
\begin{minipage}{14cm}
\begin{tabular}{lccccc}
\hline
\hline
Observation ID & Modified Julian Day    & Peak Flux\footnote{The peak flux is calculated without subtracting persistent flux.} at    &Persistent Flux at          & Persistent Flux at   & Duration  \footnote{We define the duration of bursts as the time when the flux is larger than 10 \% peak flux.}\\
               &                        &  2--10 keV (cts/s/pcu)   &  2--10 keV (cts/s/pcu)      & 30--50 keV (cts/s/pcu) &   (s)
\\ \hline
30054-04-02-01  &  50971.23  &  2051.61 $\pm$ 20.81  &  65.87 $\pm$ 0.38 &  1.48 $\pm$ 0.12  &  116 \\
30054-04-02-02  &  50972.18  &  2024.11 $\pm$ 20.67  &  66.91 $\pm$ 0.39 &  1.44 $\pm$ 0.12  &  115 \\
30054-04-03-02  &  50976.42  &  2026.33 $\pm$ 23.14  &  69.02 $\pm$ 0.44 &  1.49 $\pm$ 0.14  &  116 \\
30060-03-01-01  &  50988.83  &  1996.79 $\pm$ 22.98  &  66.12 $\pm$ 0.43 &  1.62 $\pm$ 0.14  &  117 \\
50035-01-01-01  &  51724.88  &  1939.21 $\pm$ 22.67  &  89.76 $\pm$ 0.50 &  2.46 $\pm$ 0.16  &  122 \\
50035-01-02-00  &  51725.71  &  2022.12 $\pm$ 23.06  &  91.28 $\pm$ 0.50 &  1.58 $\pm$ 0.14  &  123 \\
50035-01-02-00  &  51725.88  &  2004.82 $\pm$ 23.05  &  92.43 $\pm$ 0.51 &  2.63 $\pm$ 0.16  &  126 \\
50035-01-02-02  &  51726.72  &  1970.52 $\pm$ 22.72  &  92.78 $\pm$ 0.50 &  1.57 $\pm$ 0.14  &  126 \\
50035-01-02-04  &  51728.77  &  1976.26 $\pm$ 22.78  &  92.41 $\pm$ 0.50 &  1.53 $\pm$ 0.14  &  125 \\
50035-01-03-00  &  51811.75  &  2030.83 $\pm$ 26.69  &  91.32 $\pm$ 0.58 &  1.61 $\pm$ 0.16  &  123 \\
50035-01-03-08  &  51813.49  &  1895.19 $\pm$ 22.26  &  90.54 $\pm$ 0.50 &  1.45 $\pm$ 0.14  &  128 \\
50035-01-03-09  &  51814.01  &  2018.17 $\pm$ 26.79  &  92.26 $\pm$ 0.59 &  1.30 $\pm$ 0.17  &  130 \\
50035-01-03-10  &  51814.35  &  2039.81 $\pm$ 32.70  &  95.46 $\pm$ 0.73 &  1.51 $\pm$ 0.21  &  127 \\
70044-01-01-00  &  52484.57  &  1998.33 $\pm$ 26.39  &  104.70 $\pm$ 0.61 &  1.77 $\pm$ 0.16  &  123 \\
70044-01-01-02  &  52485.01  &  1961.22 $\pm$ 26.21  &  106.85 $\pm$ 0.62 &  1.84 $\pm$ 0.16  &  136 \\
80048-01-01-00  &  52736.53  &  1890.84 $\pm$ 25.75  &  96.70 $\pm$ 0.60 &  1.36 $\pm$ 0.17  &  123 \\
80048-01-01-04  &  52738.48  &  1904.97 $\pm$ 25.87  &  94.46 $\pm$ 0.59 &  0.98 $\pm$ 0.16  &  124 \\
80048-01-01-07  &  52738.61  &  1779.96 $\pm$ 21.66  &  92.98 $\pm$ 0.51 &  1.27 $\pm$ 0.14  &  133 \\
80048-01-01-07  &  52738.75  &  1879.88 $\pm$ 22.19  &  94.69 $\pm$ 0.51 &  1.22 $\pm$ 0.13  &  126 \\
80049-01-01-00  &  52820.56  &  1850.85 $\pm$ 22.08  &  96.37 $\pm$ 0.51 &  1.91 $\pm$ 0.14  &  135 \\
80049-01-03-02  &  52834.39  &  1915.61 $\pm$ 26.08  &  99.22 $\pm$ 0.61 &  1.51 $\pm$ 0.17  &  133 \\
80049-01-03-03  &  52835.31  &  1947.30 $\pm$ 22.65  &  98.37 $\pm$ 0.52 &  1.51 $\pm$ 0.14  &  129 \\
80049-01-02-00  &  52944.40  &  1949.54 $\pm$ 26.06  &  93.18 $\pm$ 0.58 &  1.44 $\pm$ 0.16  &  127 \\
70044-01-02-00  &  53205.20  &  1879.31 $\pm$ 25.77  &  103.15 $\pm$ 0.62 &  1.46 $\pm$ 0.17  &  131 \\
70044-01-02-01  &  53206.06  &  1913.51 $\pm$ 25.84  &  101.22 $\pm$ 0.60 &  1.79 $\pm$ 0.16  &  128 \\
90043-01-01-02  &  53206.64  &  2039.15 $\pm$ 26.69  &  99.44 $\pm$ 0.60 &  1.38 $\pm$ 0.16  &  124 \\
90043-01-01-01  &  53207.22  &  1913.53 $\pm$ 25.88  &  103.23 $\pm$ 0.61 &  1.75 $\pm$ 0.16  &  135 \\
92703-01-04-02  &  53951.25  &  1864.40 $\pm$ 44.27  &  100.58 $\pm$ 1.04 &  2.93 $\pm$ 0.29  &  135 \\
92031-01-02-00  &  53959.44  &  1472.39 $\pm$ 27.81  &  75.66 $\pm$ 0.64 &  1.43 $\pm$ 0.19  &  130 \\
90043-01-02-01  &  53962.58  &  1833.93 $\pm$ 25.33  &  104.10 $\pm$ 0.61 &  1.89 $\pm$ 0.17  &  134 \\
90043-01-02-01  &  53962.72  &  1872.21 $\pm$ 25.59  &  102.83 $\pm$ 0.61 &  1.98 $\pm$ 0.16  &  130 \\
90043-01-02-001  &  53963.28  &  1922.97 $\pm$ 44.86  &  96.81 $\pm$ 1.02 &  2.28 $\pm$ 0.28  &  128 \\
90043-01-02-001  &  53963.42  &  1746.59 $\pm$ 30.24  &  98.29 $\pm$ 0.73 &  1.87 $\pm$ 0.20  &  132 \\
90043-01-02-00  &  53963.56  &  1848.28 $\pm$ 44.01  &  98.99 $\pm$ 1.03 &  1.23 $\pm$ 0.28  &  128 \\
90043-01-03-000  &  53965.93  &  1946.33 $\pm$ 26.08  &  104.11 $\pm$ 0.61 &  1.58 $\pm$ 0.16  &  130 \\
70044-01-04-00  &  53966.37  &  1848.82 $\pm$ 25.39  &  102.15 $\pm$ 0.61 &  2.00 $\pm$ 0.16  &  138 \\
90043-01-03-01  &  53966.51  &  1815.25 $\pm$ 30.89  &  93.73 $\pm$ 0.71 &  1.99 $\pm$ 0.21  &  129 \\
90043-01-03-01  &  53966.66  &  1912.50 $\pm$ 44.77  &  97.85 $\pm$ 1.02 &  1.55 $\pm$ 0.28  &  129 \\
91017-01-01-05  &  54167.58  &  1846.81 $\pm$ 31.19  &  94.73 $\pm$ 0.71 &  1.78 $\pm$ 0.19  &  125 \\
91017-01-02-03  &  54168.32  &  1879.08 $\pm$ 31.47  &  94.75 $\pm$ 0.71 &  1.67 $\pm$ 0.19  &  134 \\
91017-01-02-11  &  54168.46  &  1927.11 $\pm$ 31.75  &  104.66 $\pm$ 0.75 &  1.64 $\pm$ 0.19  &  134 \\
91017-01-02-01  &  54169.05  &  1824.74 $\pm$ 25.23  &  102.31 $\pm$ 0.61 &  1.88 $\pm$ 0.17  &  135 \\
91017-01-02-09  &  54169.34  &  2032.24 $\pm$ 32.70  &  105.68 $\pm$ 0.76 &  1.50 $\pm$ 0.20  &  129\\

\hline
\end{tabular}
\label{table1}
\end{minipage}
\end{table*}

We extracted the persistent emission several tens of seconds prior to the bursts, assuming that the accretion rate is unchanged, and fitted the net burst spectrum with the empirical model {\it wabs} \citep{Morrison1983} $\times$ {\it blackbody}. In order to reduce the errors, we defined for each burst its average temperature as the one derived from the data within two seconds around the peak flux. The mean temperature of the 43 bursts is 2.18$\pm$0.07 keV. This result is consistent with the works by \citet{intZand1999, Ubertini1999b,Galloway2008}. After convolving with the response matrix, a spectrum with this temperature would lead to a contribution of at most $0.24 \pm 0.08$ cts/s in the energy band of 30--50 keV, which is much smaller than the corresponding average of the persistent emission, resulting in $1.69\pm 0.02$ cts/s. Therefore, the energy band  of 30--50 keV  is appropriate for investigating the burst influence upon the hard X-ray emission.

The Event-mode data was used to produce lightcurves with bin size of 1 second. Following {\it RXTE}'s recommended criterion, the background was estimated by {\it pacbackest} using the latest bright source model, and subtracted off by the {\it FTOOLS} command {\it lcmath}.\footnote{Note that the background could only be extracted from the data of Standard~2 mode with the bin size of 16 s and employed for Event-mode data.
This is a standard procedure, the details are available at http://heasarc.gsfc.nasa.gov/docs/xte/abc/contents.html}
The properties of these bursts are listed in Table~\ref{table1}, and {the distributions of their persistent and peak fluxes are shown in Figure~\ref{peak_and_persistent}.}
Table ~\ref{table1} shows that throughout the whole {\it RXTE} era, the bursts were very similar to each other.
{The persistent flux was estimated as the averaged count rate within the range of 150 to 50 s prior to each burst, and their averages were 94.79$\pm$0.09 and 1.69$\pm$0.02 cts/s at 2--10 and 30--50 keV, respectively.}
This result is consistent with the reports of \citet{Cocchi2010} where GS 1826--238 was mostly observed  in the hard state with a rather stable spectrum. Hence, we summed up all the bursts in 2--10 keV and 30--50 keV, respectively, in bin sizes of 1s and 4s.
In order to do this, we took the peak flux as a { {time}} reference for each burst and average the count rate in each bin of the individual lightcurves.
The combined lightcurve is shown in Figure~\ref{total}. There is a significant shortage in the 30--50 keV band along with the burst evolution. If we take the null hypothesis that the hard X-ray shortage derives form statistical fluctuation, a $\chi^{2}$ test, assuming the lightcurve can be fitted by a constant line,  results in a value of 508.98 under 28 dofs, implying the null hypothesis is unacceptable. The integral flux for the shortage is -58.25 $\pm$ 2.76, suggesting a significant level of $\sim$ 21$\sigma$.
This phenomenon is similar to that reported for IGR J17473-2721, 4U 1636-536 and Aql X-1, but at a higher significant level \citep{Chen2012,Ji2013,Chen2013,Maccarone2003}.

\begin{figure}[t]
\centering
\includegraphics[width=3.2in]{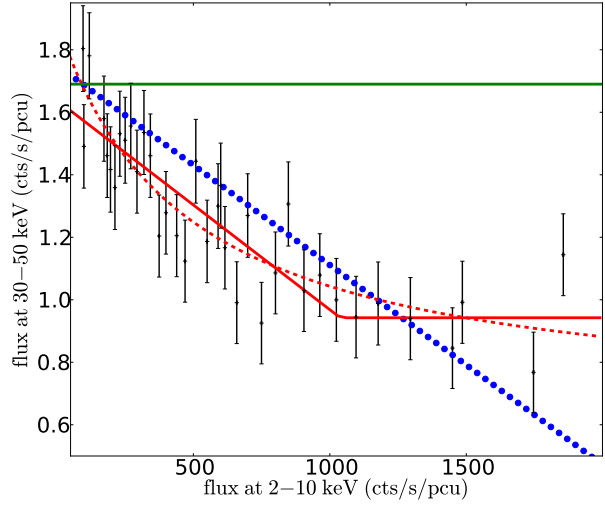}
\caption{Flux-flux diagram corresponding to the energies of 2--10 keV and 30--50 keV.
The green horizontal line represents the initial hard X-ray intensity of 1.69 cts/s.
The blue point line, the red solid line, and the red dash line show the linear fitting, piecewise fitting, and inverse proportion fitting, respectively.}
\label{flux_flux}
\end{figure}

Figure \ref{flux_flux} shows the burst flux-flux diagram. It is based on the 4~s bin lightcurves. Prior to the bursts, namely when the averaged persistent flux at 2--10 keV is 94.79$\pm$ 0.09 cts/s, the hard X-ray flux in the 30--50 keV band is {{1.69$\pm$ 0.02}} cts/s. This hard X-ray level is marked with a green line in that figure. When the bursts proceed, there is an anti-correlation between soft and hard X-ray fluxes at least until the soft count rate reaches $\sim$1000 cts/s, while the hard X-ray flattens afterwards. These results, however, have to be taken with care due to the large error bars of the hard X-ray flux measurements. To explore the significance of this behavior, we fit the data with a piecewise function ($a  x+b,x < c$; $a  c +b,x > c$, the solid line in Figure~\ref{flux_flux})
and an inverse proportion function (${a \mathord{\left/
 {\vphantom {a {(b + x) + c}}} \right.
 \kern-\nulldelimiterspace} {(b + x) + c}}$, the dash line). For the former, the resulting best fit parameters are $a= (6.67 \pm 0.66) \times {10}^{-4}$, $b=1.64 \pm 0.04$ and $c=1038.47 \pm 424.52$ with a ${\chi}^2 $ of 32.28 (34 dof); for the latter, $a= 585.45 \pm 354.54$, $b = 469.27 \pm 275.76$, and $c = 0.64 \pm 0.18$, with a ${\chi}^2 $ of 29.71 (34 dof). Both of these fits show a significant improvement
with respect to a linear fitting (shown in Figure \ref{flux_flux}).

%F-test was carried out and proved that the piecewise function could improve the fitting significantly at 5.25 $\sigma$ level.

\begin{figure}[t]
\centering
\includegraphics[width=3.2in]{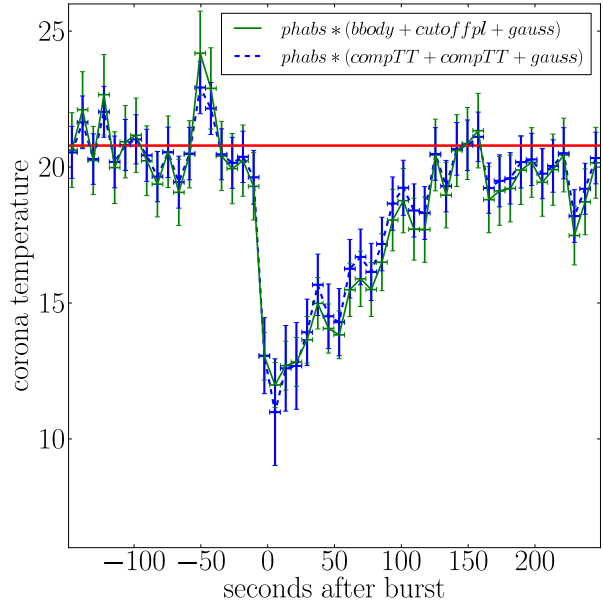}
\caption{The inferred 8-s bin corona temperature versus time around the burst as obtained using different models.
The horizontal line represents the initial temperature of $20.79$ keV.}
\label{model}
\end{figure}

Because the shape and spectrum of bursts are quite similar and the persistent flux is relatively stable, GS 1826--238 seems so far the best candidate to explore the corona behavior
under the influence of the bursts.
While bursting, the thermal component of the spectrum should increase, as well as the corresponding normalization of the compton component, because the emitted photons for the former can be regarded as the incident photons of the latter.
Meanwhile, the Compton process will take energy away from the corona and lead to a lower coronal temperature. The hard X-ray shortage we observe reveals a mixture of increasing normalization and decreasing temperature.
%Meanwhile the temperature of the corona drops rapidly.
However, limited by the quality of the spectrum, we were unable to estimate both of the variations of the normalization and temperature. Hence, we assumed that the normalization is unchanged while bursting and estimated an upper limit to the corona temperature.
In practice, based on the fitting results in Table 3 by \citet{Thompson2005}, we estimated the variation of the corona temperature by setting free $T_e$ (the temperature of the hot electrons in the corona) in {\it compTT} or $E_{\rm cut}$ in {\it cutoffpl} for which $E_{\rm cut}$ $\sim$ 2 $T_{\rm e}$. Figure ~\ref{model} shows the 8-s bin inferred corona temperature as a function of time around the bursts. The different lines in that figure represent different models used (The one is phabs*(bbody+cutoffpl+gauss), and the other one is phabs*(compTT+compTT+gauss)), the red horizontal line stands for the initial temperature of $20.79_{ - 1.08}^{ + 0.72}$ \citep{Thompson2005}.
From Figure~\ref{model} we can see that at the peak of the clock bursts the corona may have a temperature $\sim$10 keV.

To estimate the possible time lag between hard and soft X-rays, we carried out a cross correlation analysis between the lightcurves in two energy bands with 4 s bin size. In practice, to match the peaks more exactly, the data within 20 seconds around bursts (--20s, 20s) are used.
The error of the time lag was derived with Monte Carlo via sampling of the lightcurves in these two energy bands, for details see \citet{Ji2013}. We obtain a time delay of $\sim$3.6$\pm$1.2 s with the cross correlation coefficient -0.74 $\pm$ 0.03. The value of time delay is consistent in case different bin-size lightcurves are used as well.

\section{Discussion and summary}

We find a hard X-ray shortage in the  30--50 keV band promoted by the shower of soft X-ray photons coming from the type-I  bursts in GS 1826--238. This shortage happens with a time delay of 3.6 $\pm$ 1.2 seconds after the peak of the soft flux. The latter is somewhat larger and more significant than the shortages found in other sources, for example,   IGR J17473-2721, 4U 1636-536, and Aql X-1, where the observed time delays are 0.7$\pm$0.5, 2.4$\pm$1.5 and 1.8$\pm$1.5, respectively.

\subsection{The feedback of type-I  bursts to the accretion process}

{{The cooling and reheating of the corona require an energy budget.
We estimate this required energy in terms of inverse Compton scatter: ${E}_{\rm c}={E}_{\rm burst}\left({e}^{Y} -1 \right)$, %\citep{Titarchuk1994},
where ${E}_{\rm burst}$ refers to the nuclear energy released in a type-I  burst, which is $\sim 5.3 \times {10}^{39}$ ergs \citep{Galloway2004}. The {dimensionless} Y parameter \citep{Lightman1979} is given by $ \sim \frac{4k{T}_{\rm e}}{{m}_{e}{c}^{2}}Max(\tau ,{\tau}^{2})$, where ${T}_{\rm e}$ and $\tau$ are the characteristic temperature and optical depth respectively. By taking $\frac{3}{2} k{T}_{\rm e}$ and $\tau$ as 10 keV and 1, the resulting ${E}_{\rm c}$ is $\sim 3 \times {10}^{38}$ ergs, i.e, the energy of a mass of ${10}^{18}$ g accreting onto the neutron star, assuming characteristic values of 1.4 solar mass and a neutron star radius of 8 km.}
%
%If throughout the bursts the accretion rate remains constant and the required energy derives from the stored energy or mass in the inner accretion disc, we expect to observe the accretion disc moving outward by $\sim$ 50 km  and a corresponding lower luminosity, assuming a standard accretion disc and the $\alpha$ parameter is $\sim$ 0.1 \citep{Shakura1973}.
%
%
% From Frank,King, RaineÕs book, page 93, equation 5.49. The \Sigma is the surface density of the accretion disk. Additional mass of 1E18 g is needed to accrete onto the neutron star. And this value corresponds to the mass of a part of innermost disk.  int_{Rin}^{R_in+Delta} Sigma 2\pi R dR = 1E18g. If R_in=30km, then Delta=50km
%
%However, based on our observations, the flux turned out to recover to the pre-burst level along with the evolution of bursts, suggesting an averaged increase of accretion rate of at least 3\% $\dot{M}_{\rm Edd}$ during bursting.  This result shows that the feedback of bursts to the accretion process is significant on short timescales of $\sim$ 100 seconds only.
%An increased level of accretion could be the origin of the energy required for the recovery of the corona.
If throughout the burst, the accretion rate remains unchanged, the energy required for the Compton process can only
 derives from the gravitational potential energy stored by the mass in the inner accretion disc.
Assuming a standard accretion disc with $\alpha$ parameter $\sim$ 0.1 \citep{Shakura1973}, the mass in the inner accretion disk of $\sim$ 50 km should be eaten up and a corresponding lower luminosity after bursts is expected to be observed.
However, based on our observations, the flux turned out to recover to the pre-burst level along with the evolution of bursts.
Therefore, the energy should derive from an increase of accretion rate during bursting, and we estimate the increase is at least 3\% $M_{\rm edd}$.
This result shows that the feedback of bursts to the accretion process is significant on short timescales of $\sim$ 100 seconds. This result is consist with the findings of \citet{Worpel2013}.
Recently, assuming the persistent spectrum remains constant throughout the burst, they proposed a multiplicative factor $f_{\rm a}$ to the persistent emissions and took it into spectral fits while bursting. They found that the best-fit value of $f_{\rm a}$ was significantly greater than 1, suggesting the persistent flux increases during bursts.

In the accretion process, the accretion rate is limited by the efficiency of angular momentum transfer.
The increase of the accretion rate directly points to a requirement of the additional angular momentum transfer.
%
%Normally, the angular momentum transfer and loss in accretion discs depends on viscosity. During bursts, the emitted photons, which radiate along with radial direction, can interact with the accretion disk by Compton scatter. The field of scattered photons would become rotational, because the materials in accretion disk are rapidly rotational, and the scattered photons in Compton process are always along with the direction of electrons. This effect can be regarded as a force against the direction of the movement of the accreted materials and carry away their angular momentum. This effect is called Poynting-Robertson drag.
%
However,
\citet{Walker1989} proposed that the Poynting-Robertson drag could be regarded as a large torque on the inner boundary of the accretion disc and cause severe disc depletion: the inner radius of the optically thick disc moved outward. And after the burst peak, radiation-driven angular momentum transfer declines rapidly and the innermost radius of the accretion disc returns to the previous position gradually. Time-dependent numerical simulation has been shown in Figure 3 and 4 of \citet{Walker1992}, in which the estimated increase of accretion could be larger than the energy requirement for the recovery of the corona, the remaining energy may contribute to the burst luminosity.
In theory, models cannot constrain the regressive distance. In terms of the radial velocity (assuming $\alpha$ factor $\sim$ 0.1) in standard disc \citep{Shakura1973} and the characteristic burst duration of 100 seconds for hydrogen ignition and 20 seconds for helium ignition, here we estimated their upper limits as $\sim$ 60 km and $\sim$ 10 km, respectively.

%ASM observations revealed that the source intensity  slowly increased between 1997 to 2003, remaining roughly constant thereafter
%(see Figure~\ref{asm}), and during which the behavior and properties of bursts are extremely regular and quasi-periodic. Therefore, it seems impossible that the long-term increase of the accretion rate is prompted by the influence of the bursts, otherwise the flat accretion rate after 2003 should not be observed. In other words, we speculate that the influence of bursts on long-term evolution of the accretion process can not be cumulative.

\subsection{Detailed formation mechanism of the corona}

%With respect to the detailed formation mechanism, two series of models are presented: magnetic reconnection models \citep{Zhang2000,Mayer2007,Liu2002} and evaporation models \citep{Meyer1994,Esin1997,Liu2007,Frank2002}.

Based on the observed time lag of seconds between hard and soft X-rays in IGR J17473-2721, \citet{Chen2012} concluded that typical evaporation models would not be favored, because their inferred time delay, according to the viscous timescale \citep{Frank2002}, should be days, which corresponds to complete evaporation of the innermost accretion disc. However,
if we consider the influence of the additional angular momentum loss, the estimated time delay should be modified.
In classical evaporation models, the coronal evaporation results from the mechanism of "siphon-flow" (for details, see \citet{Meyer1994}).
The hot corona above the accretion disc conducts heat downward by electron conduction.
Along with the decreasing of the temperature from a high coronal value of $\sim 10^{9}$ K \citep{Narayan1995} to a low disc level of  $\sim 10^{7}$ K, heat conduction rate, which is $\propto T^{2.5}$ \citep{Frank2002}, becomes ineffective.
To radiate away the heat, the efficiency of radiation, which depends on the particle number density, should be increasing.
As a result, a corona of a given heating rate (or an accretion rate) will "dig" itself into the accretion disc and drain mass from the disc towards the corona. In this process, the key limitation is just the accretion rate, i.e., the rate at which it is able to remove the angular momentum. Therefore, it could be an overestimation to calculate the expected time delay while bursting only as the viscous timescale without considering the angular momentum loss by Poynting-Robertson drag.
If
to estimate the intrinsic timescale for the "siphon-flow" mechanism the thermal timescale for re-adjustment to thermal equilibrium is used, we obtain $\sim$ 0.03~s under typical parameters of standard disc \citep{Frank2002}.
Hence, reasonable enhancements of the evaporation models can not be ruled out as a possible formation mechanism of the corona.

Alternatively,  magnetic reconnection models can provide a self-consistent interpretation for the cooling and reheating of the corona.
In this scenario, the accretion discs around compact stars are considered in many cases to be threaded by large-scale magnetic field, which may arise from dynamo activities in the disc \citep{Pariev2007}.
{This large-scale field may form magnetic loops anchored at different radii of the disc and extending the magnetic lines into the corona \citep{Liu2002}.
Differential rotation of the disc can open the magnetic loops and magnetic reconnection can heat up the corona \citep{Dyda2013}.
Hence, during bursts, while the innermost accretion disc has been depleted \citep{Walker1992}, the magnetic reconnection should stop naturally, to be gradually enhanced again along with the inward movement of the accretion disc.
}

\acknowledgments

We acknowledge support from 973 program 2009CB824800 and the Chinese NSFC  11073021, 11133002, 11103020 and XTP project XDA04060604. DFT work is done in the framework of the grants AYA2012-39303 and SGR2009-811, and iLINK2011-0303, and was supported by a Friedrich Wilhelm Bessel Award of the Alexander von Humboldt Foundation. This research has made use of data obtained from the High Energy Astrophysics Science Archive Research Center (HEASARC), provided by NASA Goddard Space Flight Center.

\mbox{}

{}
\end{document}